\documentclass[aps,prb,twocolumn,amsmath,amssymb,superscriptaddress,showpacs]{revtex4-1}

\usepackage{graphicx}

\begin{document}

\title{Parameters for Cold Collisions of Lithium and Caesium Atoms}

\author{M.J. Jamieson}
\affiliation{Department of Computing Science, University of Glasgow, 17 Lilybank Gardens, Glasgow G12 8QQ, UK}
\email{mjj@dcs.gla.ac.uk}
\author{H. Ouerdane}
\affiliation{Mediterranean Institute of Fundamental Physics, Via Appia Nuova 31, 00040 Marino, Rome, Italy}
\affiliation{CNRT Mat\'eriaux UMS CNRS 3318, 6 Boulevard Mar\'echal Juin, 14050 Caen Cedex, France}


\begin{abstract}
We calculate the s-wave scattering length and effective range and the p-wave scattering volume for $^7$Li atoms interacting with $^{133}$Cs atoms via the X$\!^1\Sigma^+_g$ molecular potential. The length and volume are found by fitting the log-derivative of the zero energy wave function evaluated at short range to a long range expression that accounts for the leading van der Waals dispersion potential and then incorporating the remaining long range dispersion contributions to first order. The effective range is evaluated from a quadrature formula. The calculated parameters are checked from the zero energy limits of the scattering phase shifts. We comment on ill-conditioning in the calculated s-wave scattering length.
\end{abstract}

\pacs{03.65.Nk, 34.10.+x, 34.20.Cf}

\maketitle

\section{Introduction}

Du \emph{et al}. \cite{ref1} presented the s-wave scattering length for $^7$Li atoms interacting with $^133$Cs atoms via the $X^1\Sigma^+$ molecular potential. The elastic cross at very low energies, needed to interpret the behaviour of an ultra-cold ensemble of atoms, is determined not only by the s-wave scattering length $a_0$ but also by the s-wave effective range $r_{\rm e}$ and the p-wave scattering volume $a_1$.\cite{ref2} Scattering at higher angular momenta contributes little to the cross section. We calculated the length, range and volume. 

Du \emph{et al}. \cite{ref1} used the interaction potential of Staanum \emph{et al}.\cite{ref3} We consider that generally such an interaction potential $V(R)$ consists of a short range
part, consisting of a table of \emph{ab initio} values at discrete values of the atomic separation $R$ and an analytic expression that includes exchange and dispersion, and a long range dispersion tail,

\begin{equation}\label{eq1}
V_L(R) = -\frac{C_6}{R^6} - \frac{C_8}{R^8} -\frac{C_{10}}{R^{10}} 
\end{equation}

\noindent applicable from a separation beyond which exchange is negligible. In Eq. \eqref{eq1} $C_n$ denotes a van der Waals coefficient; we regard only the first three dispersion terms as being important.

With $\phi_l$ denoting $R$ times the radial parts of the zero-energy s-wave and p-wave functions with $l=0,1$ respectively ($l\leq 1$ for this potential), we find

\begin{equation}\label{eq2}
\frac{{\rm d}^2\phi_l(R)}{{\rm d}R^2} - \left[\frac{2\mu}{\hbar^2}V(R) + \frac{l(l+1)}{R^2}\right] \phi_l(R) = 0
\end{equation} 

\noindent where $\mu$ is the reduced mass of the colliding atoms and $\hbar$ is Planck's constant divided by $2\pi$. The asymptotic form of $\phi_l(R)$ depends on the length and the volume for $l=0,1$ respectively; with suitable normalisation

\begin{equation}\label{eq3}
\phi_l(R) \longrightarrow \left(\frac{R^{2l+1}}{2l+1} - a_l\right) \div R ~~~~\mbox{as}~ R\longrightarrow \infty
\end{equation}

The volume ($l=1$) is sometimes defined witout the factor three in Eq.~\eqref{eq3}.\cite{ref4} We include the leading dispersion term $-C_6R^{-6}$ exactly and use the normalisation \eqref{eq3} to obtain a form of $\phi_l(R)$ that is valid at large $R$:\cite{ref4}

\begin{equation}\label{eq4}
\phi_l^{\rm L}(R) = \sqrt{\frac{\pi \bar{a_l}R}{2}} \left[J_{-\frac{2l+1}{4}}(x) - \frac{a_l}{\bar{a_l}\sqrt{2}} J_{\frac{2l+1}{4}}(x)\right]
\end{equation} 

\noindent where $\bar{a_0} = 2\pi\gamma / \left[\Gamma(1/4)\right]^2$, $\bar{a_1} = \gamma^3\left[\Gamma(1/4)\right]^2/36\pi$, $\Gamma$ is the Gamma function, $J_{\pm\frac{2l+1}{4}}$ are Bessel functions and $x = \gamma^2/2R^2$ with $\gamma = \sqrt[4]{2\mu C_6/\hbar^2}$. The quantities $\bar{a_0}$ and $\bar{a_1}$ have dimensions of length and volume respectively and provide scaling for $a_0$ and $a_1$; $\bar{a_1} = 1.0642 \bar{a_0}^3$ is very close to the cube of $\bar{a_0}$. The log-derivative of $\phi_l^{\rm L}(R)$ is

\begin{equation}\label{eq5}
u_l^{\rm L}(R) = \frac{1}{2R} - \frac{2x}{R} \left[\frac{J'_{-\frac{2l+1}{4}}(x) - \frac{a_l}{\bar{a_l}\sqrt{2}} J'_{\frac{2l+1}{4}}(x)}{J_{-\frac{2l+1}{4}}(x) - \frac{a_l}{\bar{a_l}\sqrt{2}} J_{\frac{2l+1}{4}}(x)}\right]
\end{equation} 

\noindent where the primes indicate differentiation of the Bessel functions with respect to the argument $x$.

We calculated each scattering parameter $a_l$ as follows. Using a recurrence relation\cite{ref5} we propagated the log-derivative $u_l(R)$ of the function $\phi_l(R)$ at short range out to a separation $R^*$ beyond which exchange is negligible; $R^*\approx 30$ bohr in the present calculation. At long range we ignored the dispersion contribution $-C_8R^{-8} - C_{10}R^{-10}$ to the potential. By matching the calculated short range log-derivative $u_l(R^*)$, which reflects all the potential, to the log-derivative $u_l^{\rm L}(R)$ we obtained a value $\tilde{a}_l(R^*)$ of $a_l$ which is appropriate to the long range potential $-C_6R^{-6}$; the dependence of $\tilde{a}_l(R^*)$ on $R^*$ is a consequence only of the extent of the short range propagation with the complete potential. Having thus calculated $\tilde{a}_l(R^*)$ we found $a_l$ from 

\begin{equation}\label{eq6}
a_l = \tilde{a}_l(R^*) + \delta_l(R^*)
\end{equation} 

\noindent where 

\begin{equation}\label{eq7}
\delta_l(R^*) = -\frac{2\mu}{\hbar^2}\int_{R^*}^\infty \phi_l(R) \left[\frac{C_8}{R^8} + \frac{C_{10}}{R^{10}}\right]\phi_l(R){\rm d}R 
\end{equation}

\noindent accounts for the dispersion terms $-C_8R^{-8}$ and $-C_{10}R^{-10}$. We calculated the Bessel functions and their derivatives by the method described by Press et al.\cite{ref6} We repeated the calculation with several values of $R^*$ to obtain convergence. This method allows calculations of $a_0$ and $a_1$ without the need to solve Eq.~\eqref{eq2} at large separations. 

Equation \eqref{eq7} provides corrections to the length to first order in the dispersion contribution $-C_8R^{-8}$ and $-C_{10}R^{-10}$ with $l=0$.\cite{ref7} By substituting $Z = R^3/3$ when $l=1$ we find that the volume is equivalent to the length for a new potential $R^{-4} \times V(R)$ and a new zero-energy wave function $R\times\phi_1(R)$.\cite{ref8} Therefore the volume correction is determined by the length correction with the new potential. By substituting $R^{-4} \times V(R)$ for $V(R)$, $R\times\phi_1(R)$ for $\phi_0(R)$, $Z$ for $R^3/3$ and ${\rm d}Z$ for $R^2{\rm d}R$ in Eq. \eqref{eq7} and then recasting the new quadrature in terms of $R$ we obtain an equation for $\delta_1(R^*)$ that is identical to Eq. \eqref{eq7} for $l=1$. Hence Eq. \eqref{eq7} provides corrections for both length and volume. 

We found the range from:\cite{ref9}

\begin{equation} \label{eq8}
\frac{r_{\rm e}}{\bar{a_0}} = \frac{1}{3}\left[\frac{\Gamma(\frac{1}{4})}{\Gamma(\frac{3}{4})}\right]^2 \left[1 - 2\frac{\bar{a_0}}{a_0} + 2\left(\frac{\bar{a_0}}{a_0}\right)^2\right]
\end{equation}

\noindent This quadrature based formula does not depend on the semiclassical approximation but follows the assumption that $1/R \times \phi_l^{\rm L}(R)$ is a good approximation to the zero-energy radial wave function throughout. This depends on two assumptions about the range: there is little influence at small separations which was demonstrated semiclassically\cite{ref9} but can
also be demonstrated quantum mechanically; and the dispersion terms $-C_8R^{-8}$ and $-C_{10}R^{-10}$ have little effect, which was demonstrated by perturbation calculations.\cite{ref10} Thus the range for a pair of atoms depends almost solely on its scattering length $a_0$ and mean scattering length $\bar{a_0}$.\cite{ref9,ref10}

We, like Du et al.,\cite{ref1} used the potential of Staanum et al.\cite{ref3} The values of $\bar{a_0}$ and $\bar{a_1}$ are 44.4028 bohr and 9.31722 $\times 10^4$ bohr$^3$, respectively. We show the convergence of the length and volume in Table 1. With the approximation\cite{ref6} $J_{\nu}(x)\approx \sqrt{2/\pi x}~\cos(x-\nu\pi/2-\pi/4)$, we find:

\begin{equation}\label{eq9}
\delta_l(R^*) \approx \delta^8_l(R^*) + \delta^{10}_l(R^*)
\end{equation}

\noindent where 

\begin{eqnarray}\label{eq10}
\nonumber
\delta^8_l(R^*) & = & \frac{2\bar{a_l}C_8}{\gamma^2C_6}\sum_{i=-1}^{i=1} \left(\frac{a_l}{\sqrt{2}\bar{a_l}}\right)^{i+1}(-2^{-|i|})\\ 
&\times& \left({x^*}^2\cos\theta -x^*\cos\phi + \frac{\sin\phi - \sin\chi}{2}\right) 
\end{eqnarray}

\noindent and 

\begin{eqnarray}\label{eq11}
&&\delta^{10}_l(R^*) = \frac{2\bar{a_l}C_{10}}{\gamma^4C_6}\sum_{i=-1}^{i=1} \left(\frac{a_l}{\sqrt{2}\bar{a_l}}\right)^{i+1}(-2^{|i|})\\ 
\nonumber
&\times& \left(\frac{{x^*}^3}{3}\cos\theta - \frac{{x^*}^2}{2}\cos\phi +\frac{x^*}{2}\sin\phi + \frac{\cos\phi - \cos\chi}{4}\right) 
\end{eqnarray}

\noindent in which $x^* = \gamma^2/2{R^*}^2$, $\theta = (1-|i|)\nu\pi$, $\phi=2x^*+i\nu\pi$ and $\chi=i\nu\pi$, with $\nu=(2l+1)/4$. Evaluation of expressions \eqref{eq10} and \eqref{eq11} is rapid but the values of the approximate expression \eqref{eq9} converge less quickly than the accurate values as can be seen in Table 1; the departure from monotonic variation of the corrected
lengths and volumes near 60 bohr is a consequence of the increasing inaccuracy of the trigonometric approximation with decreasing $x$ corresponding to increasing $R$. However the approximate expression provides a useful check on our results. 

Our final values of the length, volume and range are 48.54 bohr, 1.617 $\times 10^4$ bohr$^3$ and 109.4 bohr, respectively. Our length differs by 4\% from that, 50.5 bohr, calculated by Du \emph{et al.}\cite{ref1} When the scattering length is large it is known that its calculated value can be sensitive to computational details and can even manifest spurious sign changes,\cite{ref3} but possible ill-conditioning in evaluations of various smaller scattering lengths is less well appreciated. Our reduced mass, 1.21481$\times 10^4$ atomic units obtained from the atomic masses,\cite{ref11} is essentially the same as that used by Du \emph{et al.}\cite{ref1} but we found that our calculated $^7$Li--$^{133}$Cs length varied by about 8\% when the potential was changed by only 0.1\%; a fractional reduction of 0.05\% (making the potential less negative beyond the classical turning point $R_0$) produced a length of 50.5 bohr. Thus it is plausible that our result and that of Du \emph{et al}.\cite{ref1} differ because of possible different interpolations used to calculate the potential. With the nuclear masses we found the scattering length to be 49.44 bohr. 

The cause of the sensitivity of the calculated length can be seen in the semiclassical expression for it\cite{ref9}

\begin{equation}\label{eq12}
a_0^{\rm SC} = \bar{a_0} \left[1 - \tan(\Phi - \pi/8)\right]
\end{equation}

\noindent where $\hbar\Phi$ is the action integral:

\begin{equation}\label{eq13}
\hbar\Phi = \int_{R_0}^\infty \sqrt{-2\mu V(R)}~{\rm d}R
\end{equation}

The angle $\Phi - \pi/8$ has value 172.702. The potential supports 55 bound states and the argument of the tangent function in Eq. \eqref{eq12} is the same as $\Phi - \pi/8 - 55\pi = -0.085$. A 0.1\% change in the potential alters the action integral by 0.05\% and changes $\Phi$ by 0.86 which alters $\Phi - \pi/8 - 55\pi$ by 10\%. The semiclassical length is 48.20; the value of $\Phi - \pi/8 -55\pi$ that allows Eq. \eqref{eq12} to reproduce the quantal length is −0.093 for which $\Phi - \pi/8 - 55\pi$ changes by 9\% which is consistent with our numerical prediction.

We made a simple check on our calculations by comparing the values of $\tilde{a_1}$ with those predicted from $\tilde{a_0}$ by Gao's formula\cite{ref12} 

\begin{equation}\label{eq14}
\tilde{a_1} = \bar{a_1} \times \frac{\tilde{a_0} - \bar{a_0}}{\bar{a_0}-\tilde{a_0}/2}
\end{equation} 

which applies to potentials as $R^{-6}$ as $R\longrightarrow\infty$ at separations where the Bessel functions of Eq. \eqref{eq5} are well represented by the trigonometric approximation and the rotational term $2/{R^*}^2$ is small. Table 1 shows good agreement between the volumes predicted by Eq. \eqref{eq14} for $R\approx$ 30 bohr, where these conditions are met, but the agreement deteriorates for larger values of $R$ as expected.

\begin{table}
\caption{\label{tab:table1}Scattering lengths (bohr) and volumes ($10^4$ bohr$^3$) for cold $^7$Li -- $^{133}$Cs collisions, for various values of separation $R^*$ (bohr).}
\begin{ruledtabular}
\begin{tabular}{cccccccc}
$R^*$ &$\tilde{a_0}$ &$a_0$\footnotemark[1] &$a_0$\footnotemark[2] &$\tilde{a_1}$ &$\tilde{a_1}$\footnotemark[3] &$a_1$\footnotemark[1] &$a_1$\footnotemark[2]\\
\hline
30& 53.68& 48.36& 48.28& 4.642& 4.921& 1.030& 1.122\\
35& 52.85& 48.37& 48.31& 4.098& 4.379& 1.199& 1.269\\
40& 50.68& 48.48& 48.37& 2.883& 3.069& 1.493& 1.601\\
45& 49.25& 48.54& 48.48& 2.056& 2.281& 1.602& 1.664\\
50& 48.78& 48.54& 48.53& 1.757& 2.037& 1.615& 1.632\\
55& 48.68& 48.54& 48.55& 1.675& 1.987& 1.617& 1.611\\
60& 48.67& 48.54& 48.56& 1.657& 1.981& 1.617& 1.601\\
65& 48.67& 48.54& 48.56& 1.655& 1.981& 1.617& 1.599\\
100& 48.59& 48.54& 48.54& 1.641& 1.941& 1.617& 1.614\\
500& 48.54& 47.54& 48.54& 1.617& 1.915& 1.617& 1.617\\
\end{tabular}
\end{ruledtabular}
\footnotetext[1]{With corrections \eqref{eq6}, \eqref{eq7}.}
\footnotetext[2]{With approximate corrections \eqref{eq6}, \eqref{eq9}, \eqref{eq10}, \eqref{eq11}.}
\footnotetext[3]{Predicted by Gao's formula \eqref{eq14}.}
\end{table}

As a check on all our calculated parameters we solved the s-wave and p-wave scattering equations over a large but finite range of $R$ at low energies and fitted the scattering parameters to the effective range expansions of the phase shifts.\cite{ref2} We extrapolated the results to zero energy but in doing so we included only even powers of the wave number which is legitimate for a truncated potential but the long range nature of the actual potential must be accounted for.\cite{ref13} We carried out this accounting by adding corrections.\cite{ref4,ref14} We found the same values for the volume and length. The range thus found accounts exactly for the dispersion terms $-C_8R^{-8}$ and $-C_{10}R^{-10}$; our value changed only slightly to 109.6 bohr, verifying that these terms have little influence.\cite{ref10} 

In conclusion, we have calculated the scattering parameters of length, volume and range for $^7$Li atoms interacting with $^{133}$Cs atoms via the X$\!^1\Sigma^+_g$ molecular potential as 48.54 bohr, 1.617 $\times10^4$ bohr$^3$ and 109.4 bohr, respectively.


\begin{thebibliography}{}
\bibitem{ref1} B. G. Du, J. F. Sun, J. C. Zhang, Y. Zhang, W. Li, and Z. L. Zhu, Chin. Phys. Lett. {\bf 25}, 3639 (2008).
\bibitem{ref2} N. F. Mott and H. S. W. Massey, \emph{The Theory of Atomic Collisions} (Oxford: Clarendon, 1965).
\bibitem{ref3} P. Staanum, A. Pashov, H. Kn\"ockel, and E. Tiemann Phys. Rev. A {\bf 75}, 042513 (2007). 
\bibitem{ref4} R. Smytkowski, J. Phys. A {\bf 28}, 7333 (1995).
\bibitem{ref5} D. E. Manolopoulos, M. J. Jamieson, and A. Pradhan, J. Comput. Phys. {\bf 105}, 169 (1993).
\bibitem{ref6} W. H. Press, B. P. Flannery, S. A. Teukolsky, and W. T. Vetterling, \emph{Numerical Recipes in FORTRAN 77. The Art of Scientific Programming}, 2nd edn, (Cambridge: Cambridge University, 1992)
\bibitem{ref7} M. J. Jamieson and A. Dalgarno, J. Phys. B {\bf 31}, L219 (1998).
\bibitem{ref8} F. Calogero, \emph{Variable Phase Approach to Potential Scattering} (New York: Academic, 1967).
\bibitem{ref9} V. V. Flambaum, G. F. Gribakin, and C. Harabati, Phys. Rev. A {\bf 59}, 1998 (1999).
\bibitem{ref10} A. C. Cord\'on and E. R. Arriola Phys. Rev. A {\bf 81}, 044701 (2010).
\bibitem{ref11} G. Audi and A. H. Wapstra, Nucl. Phys. A {\bf 565}, 1 (1993).
\bibitem{ref12} B. Gao, J. Phys. B {\bf 37}, 4273 (2004).
\bibitem{ref13} L. Spruch, T. F. O'Malley, and L. Rosenberg, Phys. Rev. Lett. {\bf 5}, 375 (1960); B. R. Levy and J. B. Keller, J. Math. Phys. {\bf 4}, 54 (1963); O. Hinckelmann and L. Spruch Phys. Rev. A {\bf 3}, 642 (1971).
\bibitem{ref14} M. J. Jamieson, H. Sarbazi-Azad, H. Ouerdane, G. H. Jeung, Y. S. Lee and W. C. Lee, J. Phys. B {\bf 36}, 1085 (2003); M. Marinescu, Phys. Rev. A {\bf 50}, 3177 (1994); T. Orlikowski, G. Staszewska, and L. Wolniewicz, Mol. Phys. {\bf 96}, 1445 (1998).
\end{thebibliography}
\end{document}